# L 363-38 b: a planet newly discovered with ESPRESSO orbiting a nearby M dwarf star

Lia F. Sartori[1,*], Christophe Lovis[2], Jean-Baptiste Delisle[2], Monika Lendl[2], Gabriele Cugno[1], Anna Boehle[1], Felix Dannert[1], Andrea Krenn[2,3], Jonas L. Gubler[1] and Sascha P. Quanz[1]

[1] ETH Zurich, Institute for Particle Physics and Astrophysics, Wolfgang-Pauli-Strasse 27, CH-8093 Zurich, Switzerland
[2] Département d'Astronomie, Université de Genève, Chemin Pegasi 51, 1290 Versoix, Switzerland
[3] Space Research Institute, Austrian Academy of Sciences, Schmiedlstr. 6, 8042 Graz, Austria

**ABSTRACT**

*Context.* Planets around stars in the solar neighbourhood will be prime targets for characterisation with upcoming large space- and ground-based facilities. Since large-scale exoplanet searches will not be feasible with such telescopes, it is crucial to use currently available data and instruments to find possible target planets before next generation facilities come online.
*Aims.* We aim at detecting new extrasolar planets around stars in the solar neighbourhood by blind radial velocity (RV) search with ESPRESSO. Our target sample consist of nearby stars ($d < 11$ pc) with little ($< 10$) or no previous RV measurements.
*Methods.* We use 31 radial velocity measurements obtained with ESPRESSO at the VLT between December 2020 and February 2022 of the nearby M dwarf star ($M_\star = 0.21$ $M_\odot$, $d = 10.23$ pc) L 363-38 to derive the orbital parameters of the newly discovered planet. In addition, we use TESS photometry and archival VLT/NaCo high contrast imaging data to put further constraints on the orbit inclination and the possible planetary system architecture around L 363-38.
*Results.* We present the detection of a new extrasolar planet orbiting the nearby M dwarf star L 363-38. L 363-38 b is a planet with minimum mass $m_p \sin(i) = 4.67 \pm 0.43$ $M_\oplus$ orbiting its star with a period P = 8.781± 0.007 d, corresponding to a semi-major axis $a = 0.048 \pm 0.006$ AU, which is well inside the inner edge of the habitable zone. We further estimate a minimum radius $r_p \sin(i) \approx 1.55 - 2.75$ $R_\oplus$ and an equilibrium temperature $T_{eq} \approx 330$K.
*Conclusions.* With this study, we further demonstrate the potential of the state-of-the-art spectrograph ESPRESSO for detecting and investigating planetary systems around nearby M dwarf stars, which are inaccessible to previous instruments such HARPS.

**Key words.** Planets and Satellites: detection, fundamental parameters – Techniques: radial velocities – Stars: individual: L 363-38

## 1. Introduction

In the last decades, the growing interest in exoplanetary science, as well as technical developments, have led to the discovery of over 5000 exoplanets orbiting around stars up to thousands of pc away from Earth[1]. However, a complete census of the planets in the solar neighbourhood ($d < 15$ pc) is still missing. Such planets will be the prime targets for characterisation with new and upcoming space- and ground-based facilities such as the *James Webb Space Telescope* (JWST, Gardner et al. 2006), the Enhanced Resolution Imager and Spectrograph (ERIS, Davies et al. 2018) at the Very Large Telescope (VLT), the Mid-infrared ELT Imager and Spectrograph (METIS, Quanz et al. 2015, Brandl et al. 2021) and the ArmazoNes high Dispersion Echelle Spectrograph (ANDES, formerly known as High Resolution Echelle Spectrograph, HIRES, Marconi et al. 2016) at the Extremely Large Telescope (ELT), and, especially, for future space missions such as the Habitable Exoplanet Observatory (HabEx, Gaudi et al. 2020), the Large UV/Optical/IR Surveyor (LUVOIR, Peterson et al. 2017, The LUVOIR Team 2019), the Large Interferometer for Exoplanets (*LIFE*, Quanz et al. 2018, Quanz et al. 2021, Konrad et al. 2021, Dannert et al. 2022, Alei et al. 2022), and the Closeby Habitable Exoplanet Survey (CHES, Ji et al. 2022), whose goal is to characterize Earth-like planets around the nearest stars. Since large-scale exoplanet searches will not be feasible with such telescopes, or at least very costly, it is crucial to use current available data and instruments to find possible target planets (or lack thereof) in the solar neighbourhood before next generation facilities come online.

To determine the current constraints on the planet population in the solar neighbourhood and to prioritize targets for these upcoming instruments/missions, we started a project to systematically search for planets and substellar companions around nearby stars ($d < 15$ pc), mostly using the high contrast imaging (HCI) and radial velocity (RV) techniques. This program consists of re-analyzing archival direct imaging data (e.g., Boehle et al. 2019) as well as proposing for new observations. In this paper we will concentrate on new radial velocity data obtained with the Echelle SPectrograph for Rocky Exoplanets and Stable Spectroscopic Observations (ESPRESSO, Pepe et al. 2021, González Hernández et al. 2018). ESPRESSO is the state-of-the-art ultra-stable high resolution spectrograph installed at the VLT, with a resolving power of $R \sim 140000$ covering the spectral range from $\sim 380$ nm to $\sim 788$ nm. At the best observing conditions, ESPRESSO is able to reach a precision at the level of $\sim 10$ cm s$^{-1}$ on the sky, which would allow to detect Earth-like planets around Sun-like stars. This is one order of magnitude better than its predecessor, the High Accuracy Radial ve-locity Planet Searcher (HARPS, Pepe et al. 2002, Mayor et al. 2003). ESPRESSO can collect light from either a single VLT Unit Telescope (UT) or up to four UTs simultaneously through the Coudé trains (Cabral et al. 2010), allowing for high-cadence RV observations. Additional information about the ESPRESSO instrument is available in the ESO user manual and documentation[2]. Since the start of operations in 2018, ESPRESSO has proved successful both at discovering new planets (Lillo-Box et al. 2021, Faria et al. 2022) as well as better constraining the properties of known planetary systems (e.g. Toledo-Padrón et al. 2020, Damasso et al. 2020, Mortier et al. 2020, Suárez Mascareño et al. 2020, Sozzetti et al. 2021, Leleu et al. 2019, 2021, Jordán et al. 2022).

---

* e-mail: lia.sartori@phys.ethz.ch
[1] Numbers based on the NASA Exoplanet Archive: https://exoplanetarchive.ipac.caltech.edu/

In the following we report the detection and characterisation of a planet orbiting the nearby M dwarf star L 363-38. This is one of the few standalone planets discoveries with ESPRESSO so far. The observations are described in Section 2. In Section 3 we report the stellar parameters and describe the performed analysis. The results are discussed in Section 4, and we conclude in Section 5.

## 2. Observations

### 2.1. Radial velocities

L 363-38 was observed using ESPRESSO at the VLT between December 12, 2020 and February 08, 2022 as part of a monitoring program of nearby ($d < 11$ pc) stars with little (<10) or no previous RV measurements (program ID 106.21BN.001 with carry over during P108, PI Anna Boehle). Throughout the two observing periods we obtained a total of 31 observations (7 during P106 and 24 during P108) with a median observing cadence of 2.1 days. Every spectrum was obtained with a 15 min integration time plus 10 min of telescope and instrument overhead, resulting in 25 min of telescope time per RV measurement. For this source, 8 previous RV measurements obtained with HARPS between August 7, 2003, and December 8, 2010, are also available. Since the HARPS observations have lower significance compared to the ESPRESSO ones, in this paper we will only discuss the analysis of the ESPRESSO radial velocities. However, we checked that these results are consistent with the ones obtained including the HARPS observations. Details about all RV observations are given in Table 1.

The radial velocity data used in this work were processed and partly analysed using the Data & Analysis Center for Exoplanets (DACE), a facility dedicated to extrasolar planets data visualisation, exchange and analysis hosted by the University of Geneva[3]. Specifically, we downloaded the RV time series using the DACE python APIs, and analysed them using the dedicated modules `kepmodel` (Delisle et al. 2016, 2020a,b), `samsam` (Haario et al. 2001, Andrieu & Thoms 2008, Delisle et al. 2018), `spleaf` (Foreman-Mackey et al. 2017, Delisle et al. 2020b, Gordon et al. 2020, Delisle et al. 2022) and `kepderiv`, which allow advanced statistical methods to model the Keplerian motion of planets and disentangle it from stellar activity signals. The RV and stellar activity indicators provided in DACE are computed by cross-correlating the calibrated spectra with stellar templates for the specific stellar class, using the publicly-available ESPRESSO Data Reduction Software[4].

[2] https://www.eso.org/sci/facilities/paranal/instruments/espresso/doc.html
[3] https://dace.unige.ch/dashboard/index.html
[4] https://www.eso.org/sci/software/pipelines/espresso/espresso-pipe-recipes.html

### 2.2. TESS photometry

L 363-38 (TIC 118585685) was observed by the Transiting Exoplanet Survey Satellite (TESS, Ricker et al. 2015) during Cycle 1 (Sector 2, August 23 - September 20, 2018) and Cycle 3 (Sector 29, August 26 - September 21, 2020). In order to investigate potential transits, we retrieved the 2-min cadence Presearch Data Conditioning (PDC) light curves (Smith et al. 2012; Stumpe et al. 2014) from the TESS Science Processing Operations Center (Jenkins et al. 2016).

### 2.3. High contrast imaging

L 363-38 was observed with the NAOS-CONICA instrument at the VLT (NaCo, Lenzen et al. 2003, Rousset et al. 2003) between December 07, 2003 and December 10, 2003 as part of a survey aimed at constraining stellar multiplicity at very low masses (program ID 072.C-0570A, PI J.-L. Beuzit). The observations were obtained with the narrow band infra-red filter NB_1.64 ($\lambda_0 = 1.644\ \mu m$, $\Delta \lambda = 0.018\ \mu m$) in imaging mode for a total observing time of 160 seconds. We downloaded the raw files from the ESO archive, and we reduced and analysed them with the state-of-the-art, direct imaging data pipeline PynPoint (Amara & Quanz 2012, Stolker et al. 2019).

## 3. Analysis

### 3.1. Stellar properties

L 363-38 (also known as LHS 1134 and GJ 3049, among others) is a high proper-motion nearby M4 star (Gaidos et al. 2014). The parallax reported in the *Gaia* Data Release 3 (DR3, Gaia Collaboration et al. 2016, 2021; Babusiaux et al. 2022), $\pi$ = 97.660±0.032 mas, corresponds to a distance $d$ = 10.232±0.008 pc. The apparent V-band magnitude is V = 11.51 mag which, assuming the bolometric correction for M-type stars from Habets & Heintze (1981) and the distance above, corresponds to a bolometric luminosity $L_\star$ = 0.013 $L_\odot$. The *Gaia* DR3 provides an effective temperature $T_{eff}$ = 3128.91 K and a surface gravity log g = 4.97 dex. Its stellar mass is $M_\star$ = 0.21 ± 0.014 $M_\odot$ (Winters et al. 2021 and references therein). Maldonado et al. (2020) provide a stellar age of 8.07 ± 4.12 Gyr. We further checked that the star is not in a binary system by comparing the significance of the excess noise in the *Gaia* DR2 (Gaia Collaboration et al. 2018; Arenou et al. 2018; Lindegren et al. 2018) astrometric fit ($D$) to the distribution of this parameter for a volume limited sample of M stars with $d < 11$ pc. Vrijmoet et al. (2020) found some perturbation in the astrometric residual which may indicate the presence of a companion, but with the available data they were not able to confirm and characterise it. We also checked that the source is not labeled as a non single star in the *Gaia* DR3. The general properties of L 363-38 are listed in Table 2.

### 3.2. Radial velocities

The RV time series from the ESPRESSO observations are shown in Fig. 1 (left). In addition, Fig. 2 shows the generalised Lomb-Scargle periodogram (Cumming et al. 2008, with formalism from Zechmeister & Kürster 2009) of the ESPRESSO RV and of some stellar activity indicators: full-width at half-maximum (FWHM) of the cross-correlation function (CCF), CCF Contrast and Bis-span, and Log($R'/HK$). The RV periodogram shows a clear peak at P = 8.781d and its 1 (sideral-)day aliases at P = 0.9 d and P = 1.1 d; no significant peaks at these positions

**Table 1.** Summary of the radial velocity observations.

| Instrument | # obs. | date first obs. | date last obs. | $\Delta t_{RV}^{(a)}$ [days] | $a$ for $\Delta t_{RV}^{(b)}$ [AU] | median $\tau_{RV}^{(c)}$ [days] | $\sigma_{RV}$ [m/s] |
|---|---|---|---|---|---|---|---|
| ESPRESSO P106 | 7 | 2020/12/12 | 2021/02/01 | 51 | 0.16 | 5.4 | 5.1 |
| ESPRESSO P108 | 24 | 2021/10/12 | 2022/02/08 | 119 | 0.28 | 1.1 | 2.9 |
| HARPS03 | 8 | 2003/08/06 | 2010/12/08 | 2681 | 2.25 | 126.7 | 6.3 |
| All | 39 | 2003/08/06 | 2021/11/27 | 6761 | 4.16 | 2.8 | 4.0 |

**Notes.** $^{(a)}$ Total time baseline of the RV measurements. $^{(b)}$ Semi-major axis for a revolution around a star with same stellar mass as L 363-38 with period equal to the RV time baseline. $^{(c)}$ Median time lag between two consecutive RV measurements.

**Table 2.** Stellar properties of L 363-38.

| Parameter | Value | Ref. |
|---|---|---|
| IDs | L 363-38, LHS 1134, GJ 3049 | |
| Gaia EDR3 ID | 4999067471248893952 | [1] |
| RA, DEC | 10.855, -41.295 | [1] |
| Parallax [mas] | 97.660 ± 0.032 | [1] |
| Distance [pc] | 10.232 ± 0.008 | [1] |
| $\mu_\alpha$ [mas/yr] | -488.487 ± 0.022 | [1] |
| $\mu_\delta$ [mas/yr] | -582.314 ± 0.024 | [1] |
| G [mag] | 11.55 | [1] |
| $(G_{B_p} - G_{R_p})$ [mag] | 1.48 | [1] |
| $L_\star$ [$L_\odot$] | 0.013 | [2] |
| V [mag] | 11.51 | [3] |
| 2MASS J [mag] | 8.572 ± 0.019 | [4] |
| 2MASS H [mag] | 8.03 ± 0.03 | [4] |
| 2MASS $K_S$ [mag] | 7.710 ± 0.016 | [4] |
| $T_{eff}$ [K] | 3128.9 | [1] |
| log g [dex] | 4.97 | [1] |
| $M_\star$ [$M_\odot$] | 0.21 ± 0.014 | [5] |
| $R_\star$ [$R_\odot$] | 0.274 | [6] |
| Stellar Age [Gyr] | 8.07 ± 4.12 | [7] |

**References.** : [1] Gaia Collaboration et al. (2021), [2] this work, [3] Simbad Astronomical Database (Wenger et al. 2000), [4] Cutri et al. (2003), [5] Winters et al. (2021) and references therein, [6] Pecaut & Mamajek (2013), [7] Maldonado et al. 2020.

are visible in the periodograms of the stellar activity indicators (Fig. 2). Moreover, the RV time series show very weak or no correlation with these and others activity indicators (|R| = 0.25 for the CCF-FWHM, and |R| < 0.15 for the other indicators), suggesting no significant contribution from stellar activity to the RV variations. We also tried a linear detrending of the RV time series with some of the activity indicators, but we did not observe any significant change in the RV periodogram. Altogether, these findings point to the presence of a companion orbiting L 363-38 with a period P = 8.781 d, which corresponds to a semi-major axis $a$ = 0.048 AU.

To better characterise the signal observed at P = 8.781 d and to investigate the possible presence of other signals (and thus other companions), we modeled the RV time series with a Keplerian model following the formalism of Delisle et al. (2016), and additional parameters taking into account the instrumental jitter and the RV offset. Since no significant peak is present in the RV periodogram after subtracting this simple model (Fig. 2), in the following analysis we assume the presence of one companion only.

We further refined the one-planet Keplerian model by running a Markov-Chain Monte-Carlo (MCMC) with 200'000 iterations following the algorithm described in Díaz et al. (2014, 2016). We defined log-uniform priors (see Table 3) and started the iteration at the values corresponding to the best fit results of the previous Keplerian fit. The posterior distributions for this model are shown in Fig. 3. Assuming the stellar mass listed in Table 2, the obtained period P and semi-amplitude K indicate the presence of a companion with minimum mass $m_p \sin(i) = 4.67 \pm 0.43$ M$_\oplus$ orbiting L 363-38 with a semi-major axis $a = 0.048 \pm 0.006$ AU. The derived orbital and planet parameters are listed in Table 3.

### 3.3. Constraints on the planetary system's architecture

We computed the mass limits from the residuals of the RV time series (after subtracting the Keplerian model for one planet) by following the method proposed by Bonfils et al. 2013 (and references therein). These are shown in Fig. 4, together with the values found for the planet and the position of the expected habitable zone (HZ)[5]. Specifically, for every period P this plot shows the smallest minimum mass $m_p\sin(i)$ an additional companion should have in order to be detected with the current RV data: everything lying in the parameter space above this curve can be discarded based on the available data.

As described in detail in Bonfils et al. 2013, to compute the mass limits we first created 1000 time series by shuffling our RV residuals, computed the correspondent periodograms, and found the maximum power of each periodogram. The power which divides the lower 99% from the upper 1% of the maximum power's sample is then assumed to represent the 1% false alarm probability (FAP). Successively, for every period $P_{per}$ probed by our original RV periodogram (Fig. 2) we simulated 12 RV time series by adding to the residuals a sinusoidal with period $P_{per}$, one of 12 equi-spaced phases T, and an initial semi-amplitude $K_{in} = 0.5$.

---

[5] The habitable zone (HZ) was computed following two methods. The first one (labeled as *method 1* in Fig. 4) based on the runaway and maximum Greenhouse limits in Kopparapu et al. (2013, 2014), and the second one (*method 2*) based on the Bolometric correction of Habets & Heintze (1981) and the scaling values for the inner and outer radius from Kasting et al. (1993), Kasting (1996) and Whitmire & Reynolds (1996).

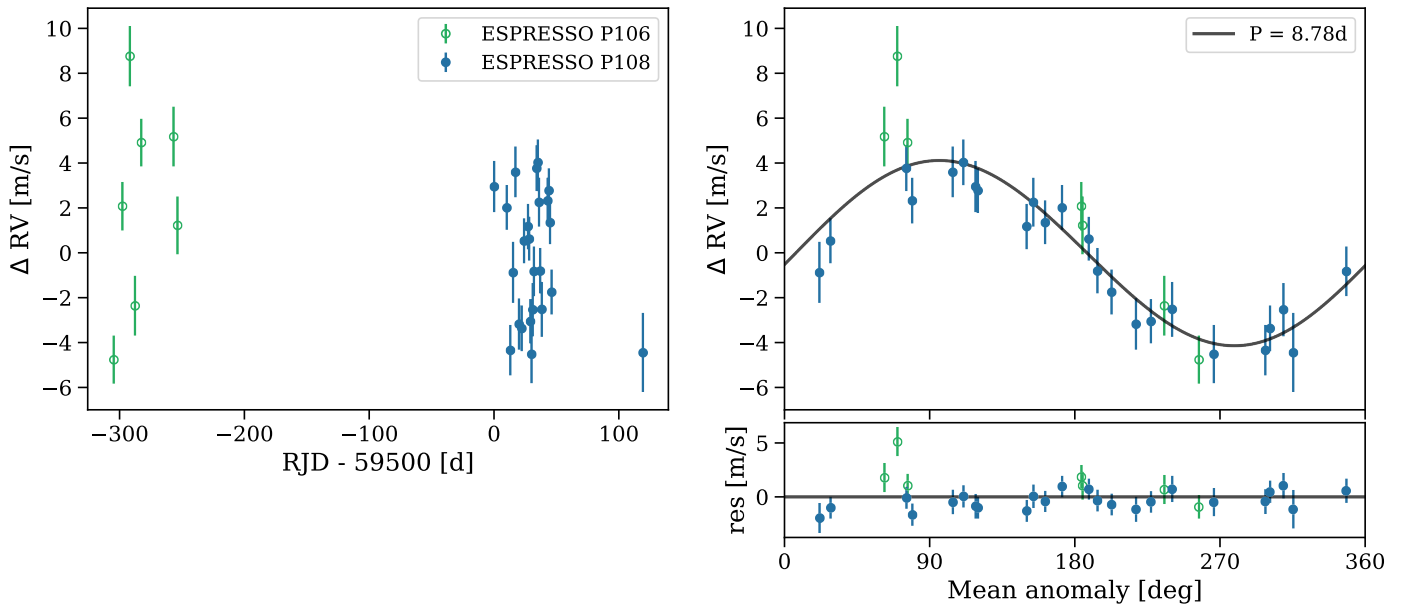

**Fig. 1.** *Left*: ESPRESSO radial velocities of L 363-38 after subtracting the linear offset and adding the instrumental jitter to the error bars. The measurements obtained during P106 and P108 are shown as open green and filled blue symbols, respectively. *Right*: Phase folded radial velocity curve of the 8.781d signal. The measurements corresponding to P106 and P108 are defined as on the left panel, while the gray curve shows the best fit Keplerian model. The bottom panel shows the residuals from the best fit model.

**Table 3.** Inferred and derived parameters for the one-planet model of the ESPRESSO radial velocities.

| Parameter | Prior | Posterior |
|---|---|---|
| *Orbital parameters* | | |
| Orbital period, $P_b$ [d] | $\mathcal{LU}(0.1, 10^5)$ | 8.781±0.007 |
| Mean longitude at t=0, $L0_b$ | $\mathcal{LU}(10^{-3}, 10)$ | 0.66±0.09 |
| RV semi-amplitude, $K_b$ [m/s] | $\mathcal{LU}(0.1, 10^5)$ | 4.12±0.33 |
| $e\cos(\omega)$ | $\mathcal{B}(0.867, 3.03)$ | -0.001±0.04 |
| $e\sin(\omega)$ | $\mathcal{B}(0.867, 3.03)$ | -0.011±0.05 |
| *Instrumental parameters* | | |
| Instrumental offset | | -27877.84±0.24 |
| Instrumental jitter | $\mathcal{LU}(10^{-3}, 10^2)$ | 0.89±0.27 |
| *Derived parameters* | | |
| Planet mass, $m_p \sin(i)$ [$M_\oplus$] | (derived) | 4.67±0.43 |
| Orbit semi-major axis, $a$ [AU] | (derived) | 0.048±0.006 |

We then increased $K_{in}$ until the correspondent signal in the periodogram reached the 1% FAP power, and converted the obtained K to $m_p \sin(i)$ using the stellar mass listed in Table 2. The mass limits shown in Fig. 4 correspond to the average over the 12 trial phases.

In addition, we computed the constraints on companion mass and semi-major axis for possible additional planets orbiting L 363-38 by following the procedure in Boehle et al. (2019). These calculations are based on the ESPRESSO residuals after subtracting the signal from the planet. Compared to the mass limits described above, these calculations allow us to add constraints for the mass of the planet itself, $m_p$, instead of the minimum mass $m_p \sin(i)$, and to investigate the presence of planets with period much larger than the RV baseline (and thus larger semi-major axis), whose signal would not be seen as periodic in the available RV data and therefore hardly be detectable in the periodogram. Specifically, the 2d histogram in Fig. 5 shows the percentage of further companions with mass and semi-major axis in a given bin which would be detected with the current ESPRESSO observations, if present. First, for every mass and semi-major axis combination in a grid with $0 < m_p < 55$ $M_\oplus$ and $0 < a < 2.5$ AU we simulated Keplerian orbits by randomly drawing 10'000 times the inclination $i$ (for randomly oriented orbital planes) and the argument of periastron $\omega$ (uniform distribution spanning $0° - 360°$). The eccentricity is set to 0 (we only consider circular orbits), the time to closest approach $T_0$ is set to a fix arbitrary number since it is degenerate with $\omega$, and the longitude of ascending node $\Omega$ is set to 0 since it does not affect the RV signal. For each of the 10 000 random orbits we then computed the expected RV signal at the epochs of our ESPRESSO observations. If the maximum RV difference measured across the observations (i.e. $RV_{max}$ - $RV_{min}$) is higher than 5 times the standard deviation of the measured RVs, then we considered the planet signal as detectable. The completeness for each combination of mass and semi-major axis was then determined from the percentage of detectable planets in the corresponding bin. As expected, RV is mostly sensitive to planets with high mass and/or small semi-major axis. Appendix A shows a similar analysis where we combine the new ESPRESSO RV data with the archival NaCo HCI data to get additional constraints on possible companions around L 363-38.

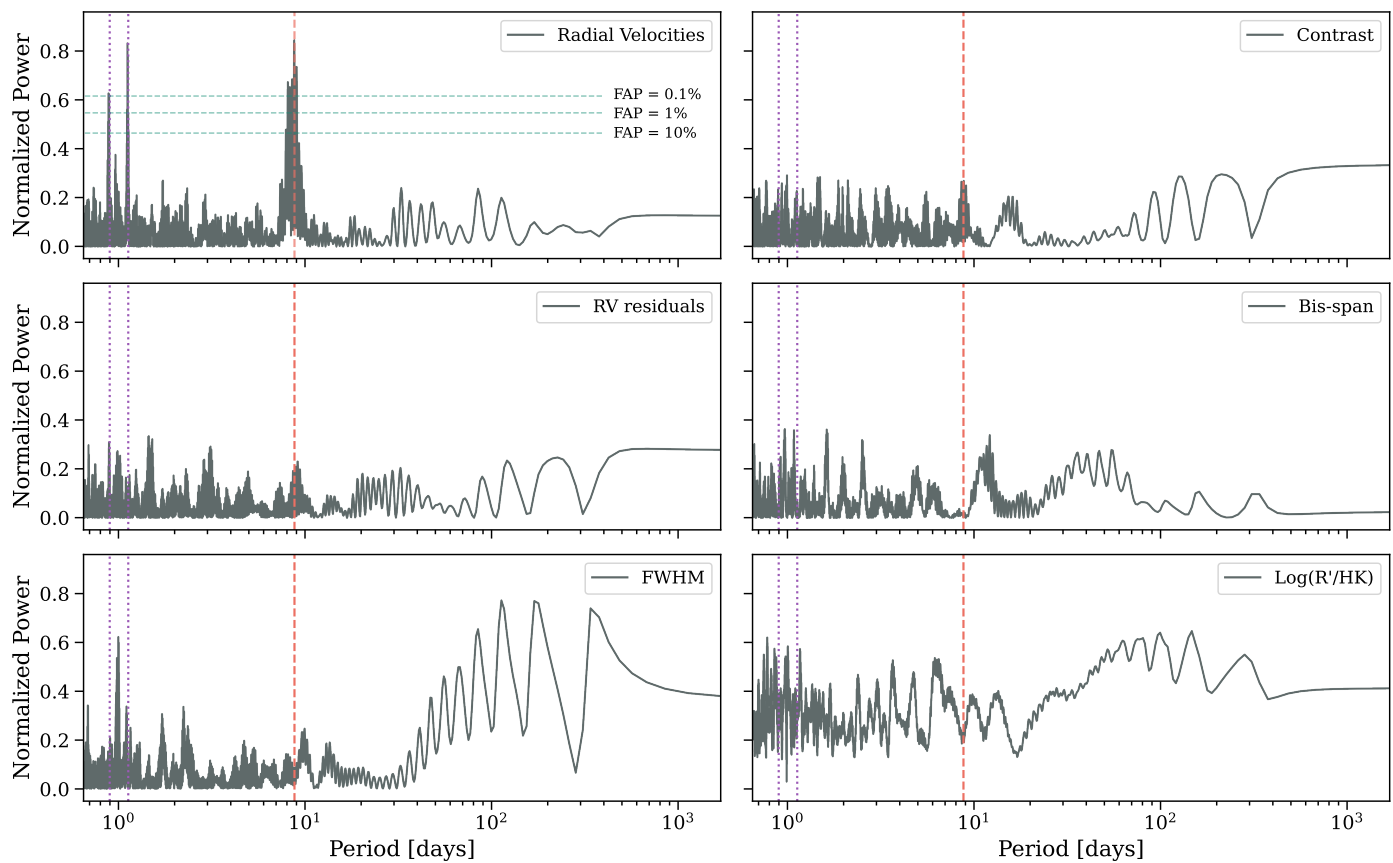

**Fig. 2.** Periodograms of the ESPRESSO time series (RV data and stellar activity indicators). The vertical lines mark the position of the companion (P = 8.781d, red dashed line) and of the aliases (violet dotted lines). A fainter peak at P = 8.781d is observed also in the Contrast indicator (although not significant), but the periodogram of the RV does not show a significant change after applying a linear detrending with any of the indicators. The 0.1%, 1% and 10% false alarm probability (FAP) levels for the RV periodogram are showed as green dashed horizontal lines.

### 3.4. TESS photometry

The TESS (Ricker et al. 2015) satellite observed L 363-38 in Sectors 2 and 29, obtaining near-continuous 2-minute cadence data throughout ∼27 days each. The Target Pixel File (TPF) for Sector 2 is shown in Fig. 6. The target is well detected, while no additional source with $\Delta m < 6$ mag is found in the same extraction aperture. Should the system be oriented edge-on, several transits would have been caught by TESS. Owing to the small host star, for which we assume here a radius of 0.274 $R_\odot$ following Pecaut & Mamajek (2013), even an unexpectedly small planetary radius of 1 $R_\oplus$ would create a transit depth of 1119 ppm.

This would be easily detectable in the TESS data, which show an RMS of 270 ppm when phase-folded on the planetary period and binned into 10-minute intervals. The duration of a transit would be expected to be at most 1.8 hours.

We therefore searched the TESS data for any transit signals that might stem from L363-38 b. To do so, we used the TESS Science Processing Operations Center (Jenkins et al. 2016) 2-min cadence PDC light curves (Smith et al. 2012; Stumpe et al. 2014), which we additionally detrended by using a boxcar filter with a width of 12 hours, much larger than any transits potentially created by L363-38 b. We computed the predicted transit windows during the TESS observations using the radial-velocity data and used the ephemeris obtained, $T_0$ = 59510.457094, $P$ = 8.557 ± 0.012 to compute the transit windows during the TESS observations[6]. Visual inspection of the raw and phase-folded data did not reveal any evident transit features at the expected times (see Fig. 7 and 8). To carry out a more robust search for potential low-amplitude transits, we used the Transit Least Squares (TLS) algorithm (Hippke & Heller 2019). As no significant features were detected we conclude that L363-38 b is not transiting.

---

[6] The period used here slightly differs from the one considered in the rest of the paper since in order to estimate the mid-transit time $T_0$ we rerun the RV analysis with a different parametrisation of the Keplerian orbits.

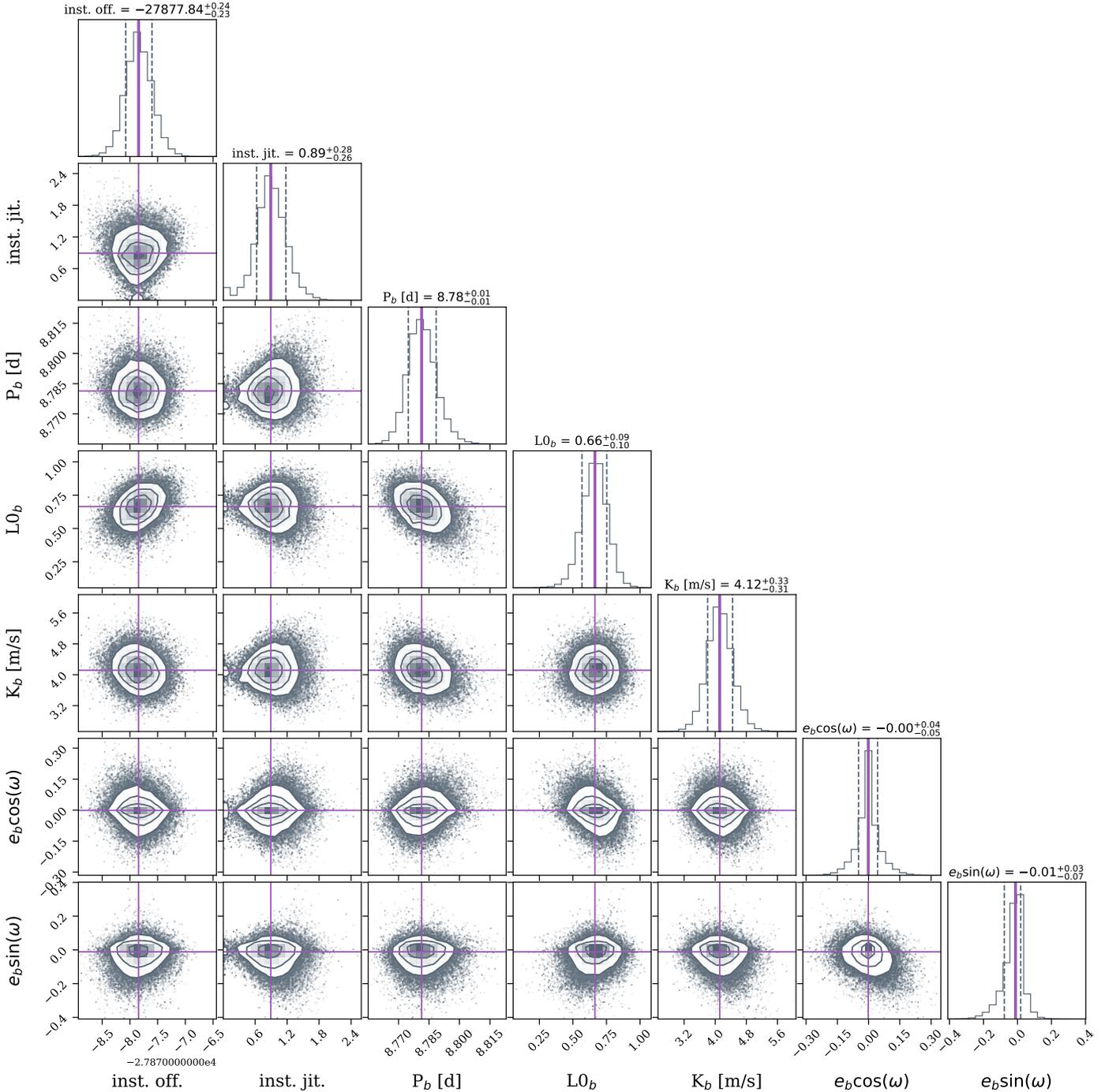

**Fig. 3.** Posterior distribution of the parameters of the Keplerian model for the ESPRESSO data, together with a linear instrumental offset and instrumental jitter. The parameters considered for the Keplerian one-planet model are period ($P_b$), mean longitude at $t = 0$ ($L0_b$), radial velocity semi-amplitude ($K_b$), eccentricity ($e_b$) and argument of periastron ($\omega$). The figure has been made using the corner python package (Foreman-Mackey 2016).

## 4. Discussion

The analysis described above points to a situation where L 363-38 is orbited by a companion with minimum mass $m_p \sin(i) = 4.67 \pm 0.43$ $M_\oplus$. Assuming the empirical mass-radius relations from Otegi et al. (2020), this minimum mass translates to a minimum radius $r_p \sin(i) = 1.61 \pm 0.06$ $R_\oplus$ for densities $\rho > 3.3$ g cm$^{-3}$ and $r_p \sin(i) = 1.85 \pm 0.3$ $R_\oplus$ for densities $\rho < 3.3$ g cm$^{-3}$, while using the Forecaster tool (Chen & Kipping 2017) we computed a minimum radius $r_p \sin(i) = 2.03^{+0.72}_{-0.68}$ $R_\oplus$. In order to overcome the planetary-mass regime[7], the orbital inclination should be $i < 0.06°$. Assuming a random orientation of the orbital planes $i$, this translates to a probability of 99.99% for the companion being a planet. Similarly, the orbital inclination should be $i > 15.8°$ in order for the planet to have a mass smaller than the mass of Nep-tune (probability 96%). To our knowledge, no additional constraint on the inclination $i$ is available to date.

---

[7] Here we consider the deuterium-burning mass limit M ∼ 13$M_J$ as a criterion to distinguish between brown dwarfs and planets (e.g. Burrows et al. 1997, Boss et al. 2007, Spiegel et al. 2011).

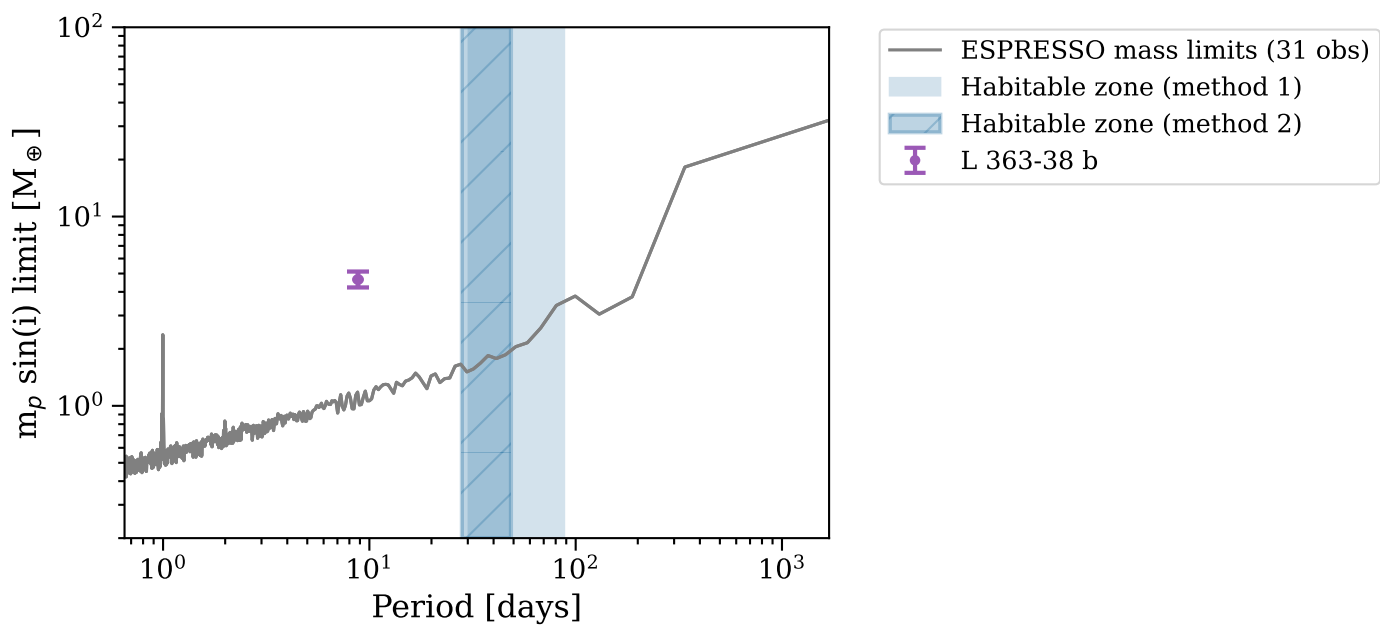

**Fig. 4.** Mass limits computed from the residuals using a bootstrap method (gray line, following Bonfils et al. 2013 and references therein), and position of the companion (purple). The blue shadowed region shows the periods corresponding to the Habitable zone (HZ) computed following two different methods. Method 1 is based on the runaway and maximum Greenhouse limits in Kopparapu et al. (2013, 2014), while method 2 relies on the Bolometric correction of Habets & Heintze (1981) and the scaling values for the inner and outer radius from Kasting et al. (1993), Kasting (1996) and Whitmire & Reynolds (1996). In both cases the planet reported here orbits at separations much shorter than the HZ.

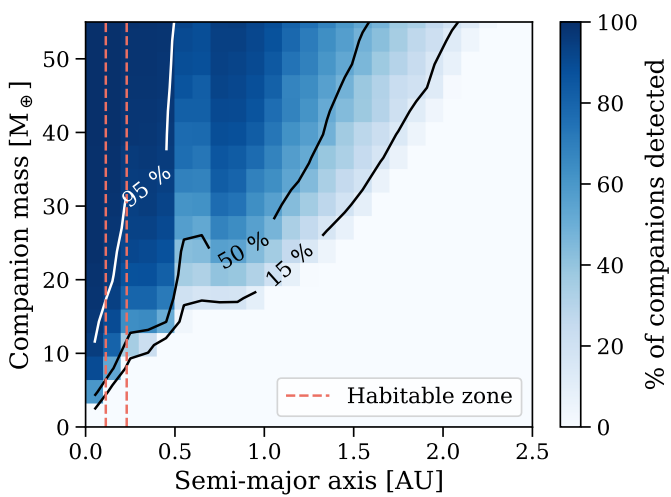

**Fig. 5.** Constraints on companion mass and semi-major axis for planets orbiting around L 363-38, based on the ESPRESSO residuals after subtracting the signal from the planet. The 2d histogram shows the percentage of companions with mass and semi-major axis in a given bin which would be detected with the current ESPRESSO observations, if present. The dashed orange vertical lines indicate the limits of the Habitable zone computed following Kopparapu et al. (2013, 2014) (method 1 in Fig. 4, see text for details).

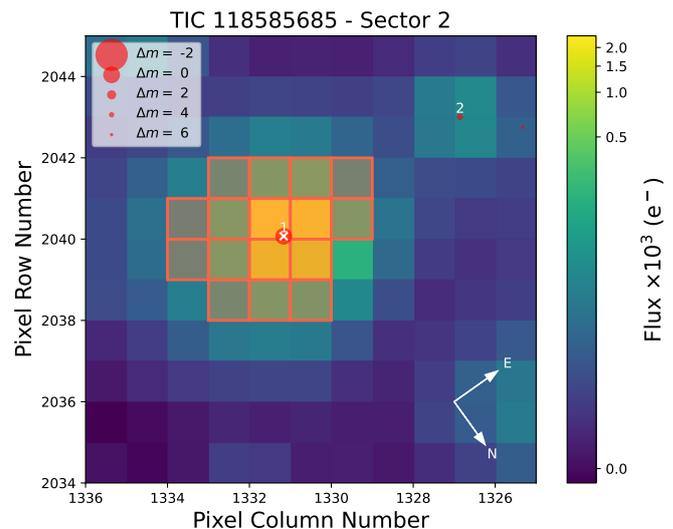

**Fig. 6.** TESS Target Pixel File (TPF) plot of L 363-38 (TIC 118585685) in Sector 2. The orange area corresponds to the TESS aperture, while the red points indicate the position of Gaia DR2 sources with magnitudes up to ∆mag = 6 compared to the target. Our target is well detected, while no additional source is found in the same aperture. The figure has been made using the `tfplotter` python package (Aller et al. 2020).

Fig. 4 shows the position of the planet in the mass-period plane together with the mass limits computed from the residuals of the ESPRESSO observations (after subtracting the Keplerian model for one planet) following the method proposed by Bonfils et al. 2013 (and references therein). We note that the planet found with this analysis is approximately a factor ~ 5 more massive than our detection limits predicted for its period. In addition, it is located well inside the inner edge of the HZ. Assuming the stellar properties listed in Table 2, the semi-major axis found with this analysis and a bond Albedo of 0.3 similar to the values observed for the Earth and Neptune (e.g. Madden & Kaltenegger 2018) we estimate an equilibrium temperature for the planet $T_{eq} \approx 330$K. According to this analysis, we can exclude the presence of planets with $m_p \sin(i) > 2 - 3\ M_\oplus$ in the HZ. Similarly, Fig. 5 suggests that only ~ 50% of the planets with $m_p \sim 10\ M_\oplus$ would be detected in the HZ, if present, while this percentage goes down to $\sim 15\%$ for $m_p \sim 5\,M_\oplus$.

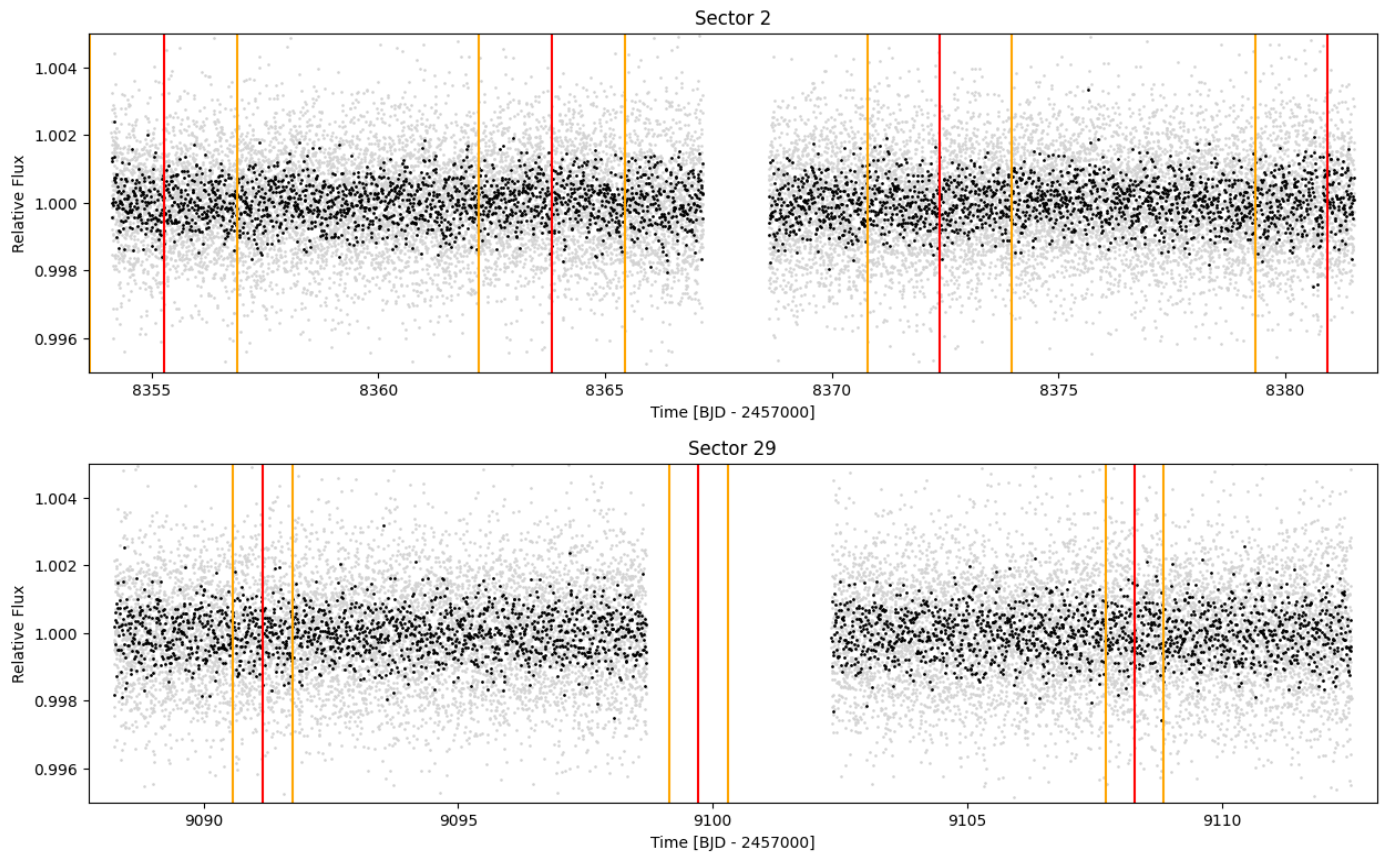

**Fig. 7.** TESS 2-minute cadence PDC SAP data of L363-38. The unbinned data are shown as grey points, and the data binned into 30-minute intervals are shown in black. The predicted times of transit and their 1-$\sigma$ probability ranges are indicated with vertical red and orange lines, respectively.

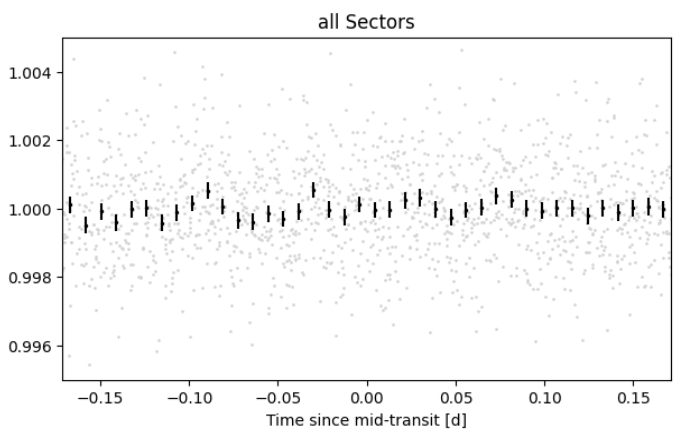

**Fig. 8.** All TESS data phase-folded on the best planetary ephemeris. The unbinned data are shown as grey points, and the data binned into 30-minute intervals are shown in black. No indication of a flux drop corresponding to a planetary transit is seen.

M dwarfs account for $\sim 75\%$ of the stars in the Milky Way, with an expected occurrence rate of Earth-sized planets (0.5-1.5) $R_\oplus$ within the HZ of $\sim 0.25 - 0.5$ planets per star, depending on the study (e.g. Kopparapu 2013, Dressing & Charbonneau 2015). However, the orbital separation of L 363-38 b is well inside the HZ, and the planet is unlikely to have liquid water on its surface and to host life. In addition, stellar emission from M dwarfs, e.g. in form of flares and winds, could highly affect the planet's atmosphere and even lead to atmospheric escape (e.g. Segura et al. 2010, Tilley et al. 2019, Kreidberg et al. 2019, Peacock et al. 2020 and references therein), and the accretion of water onto planets around M-stars during formation is diversely discussed in literature (e.g. Raymond et al. 2007, Lissauer 2007, Hansen 2015, Alibert & Benz 2017). As L 363-38b orbits its star at small separations, it is likely to be tidally locked, what would make habitability even more challenging (e.g. Barnes 2017 and references therein). Observations with new and upcoming telescopes such as JWST and METIS will be very valuable to characterise the atmosphere of L 363-38 b (or lack thereof), and to investigate these scenarios. However, since the planet is not transiting and can not be spatially resolved by neither of the two instruments, atmospheric studies will only be possible through not-transiting techniques such as e.g. phase-curves spectroscopy (e.g. Tinetti et al. 2018, Parmentier & Crossfield 2018, Glidic et al. 2022).

Fig. 9 shows a comparison between the mass, semi-major axis (*a*) and distance (*d*) distributions of the confirmed planets in the NASA Exoplanet Archive, and three planets detected with ESPRESSO: L 363-38 b (this work), HD 22469 b (Lillo-Box et al. 2021) and Proxima b (Suárez Mascareño et al. 2020). L 363-38 b and HD 22469 b have very similar parameters, while Proxima b has a lower mass and orbits the nearest stellar neighbour to the Earth. L 363-38 b is found in the lowest 8%, 18% and 5% of the $m_p \sin(i)$, $a$ and $d$ distributions of currently confirmed exoplanets, respectively.

## 5. Conclusions

We report the discovery of a planet with minimum mass $m_p \sin(i) = 4.67 \pm 0.43$ $M_\oplus$ orbiting the nearby M dwarf star L 363-38 with a period P = 8.781± 0.007 d, corresponding to a separation $a = 0.048 \pm 0.006$ AU. We further estimate a minimum radius $r_p \sin(i) \approx 1.55 - 2.75$ $R_\oplus$ and an equilibrium temperature $T_{eq} \approx 330$K. The planet was found during a blind ESPRESSO radial velocity search for planets around nearby stars, as such planets will be prime targets for characterisation with new and upcoming space- and ground-based facilities.

With this study we further demonstrated the potential of ESPRESSO for investigating planetary systems around nearby M dwarfs. Indeed, the faintness of M stars makes them challenging targets for RV studies using instruments like HARPS behind a 3.6-m telescope, but a spectrograph behind an 8-m telescope like ESPRESSO can gather sufficient light to precisely measure their RV in an efficient manner. The high precision is crucial to detect low-mass planets, as well as recognising and modeling signals arising from the stellar activity, which can easily be mistaken for planet signals (e.g. Queloz et al. 2001, Santos et al. 2014, Robertson & Mahadevan 2014, Faria et al. 2020, Suárez Mascareño et al. 2020, Lillo-Box et al. 2021). Moreover, as ESPRESSO extends further into the red part of the spectrum compared to HARPS ($\lambda_{max} \sim 790$ nm vs. $\lambda_{max} \sim 690$ nm), it is able to detect some of the flux from late M stars which is missed by HARPS.

Based on *Kepler* and TESS statistics, planets around M dwarfs are expected to occur in multi-planet systems (e.g. Dressing & Charbonneau 2015, Gaidos et al. 2016, Cloutier & Menou 2020, Hadegree-Ullman et al. 2020, Hsu et al. 2020, Feliz et al. 2021). Follow-up observations of L 363-38 with new and upcoming telescopes such as JWST and METIS, as well as astrometric data from Gaia, will be crucial to searching for additional planets further out, as well as characterising the atmosphere of L 363-38 b (or lack thereof) and improving its mass and orbital estimations.

*Acknowledgements.* This work has been carried out within the framework of the National Centre of Competence in Research PlanetS supported by the Swiss National Science Foundation. All authors acknowledge the financial support of the SNSF. ML acknowledges support of the Swiss National Science Foundation under grant number PCEFP2_194576. This publication makes use of The Data & Analysis Center for Exoplanets (DACE), which is a facility based at the University of Geneva (CH) dedicated to extrasolar planets data visualisation, exchange and analysis. DACE is a platform of the Swiss National Centre of Competence in Research (NCCR) PlanetS, federating the Swiss expertise in Exoplanet research. The DACE platform is available at https://dace.unige.ch. This work made use of tpfplotter by J. Lillo-Box (publicly available in www.github.com/jlillo/tpfplotter), which also made use of the python packages astropy, lightkurve, matplotlib and numpy. This research has made use of the SIMBAD database, operated at CDS, Strasbourg, France. This research has made use of the NASA Exoplanet Archive, which is operated by the California Institute of Technology, under contract with the National Aeronautics and Space Administration under the Exoplanet Exploration Program. This research made use of Lightkurve, a Python package for Kepler and TESS data analysis (Lightkurve Collaboration, 2018). This work has made use of data from the European Space Agency (ESA) mission *Gaia* (https://www.cosmos.esa.int/gaia), processed by the *Gaia* Data Processing and Analysis Consortium (DPAC, https://www.cosmos.esa.int/web/gaia/dpac/consortium). Funding for the DPAC has been provided by national institutions, in particular the institutions participating in the *Gaia* Multilateral Agreement.

*Authors contributions.* LAS carried out the analyses, created part of the figures, and wrote the bulk part of the manuscript. CL reprocessed the ESPRESSO data. CL and JBD supported with the analysis of the RV data. ML and AK carried out the analysis of the TESS data, and ML created part of the figures and wrote part of the manuscript. GC reduced the NaCo data, created the contrast curves, and wrote part of the manuscript. AB wrote the initial ESO proposal and some code for the data analysis. SPQ and CL initiated the project. All authors discussed the results and commented on the manuscript.

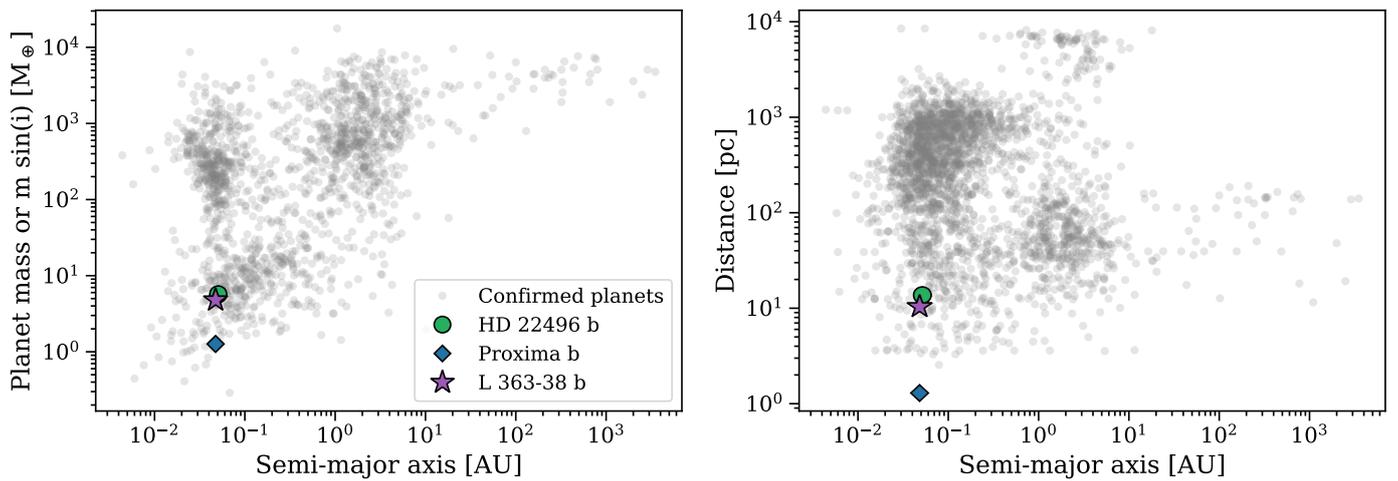

**Fig. 9.** Comparison between the mass, semi-major axis and distance distributions of the confirmed planets in the NASA Exoplanet Archive (gray shaded dots) and three planets detected with ESPRESSO: HD 22469 b (green circle, Lillo-Box et al. 2021), Proxima b (blue diamond, Suárez Mascareño et al. 2020) and L 363-38 b (purple star, this work).

## Appendix A: Combining RV and HCI constraints

As discussed in the introduction, the goal of our solar neighbourhood program is not only detect new companions, but also to put strong constraints on the possible planetary architecture using the available data and non detections. In this Appendix we follow the method described in Boehle et al. (2019) and combine the new ESPRESSO RV data with archival NaCo HCI to put constraints on possible additional companions around L 363-38.

Figure A.1 (left) shows the contrast limits obtained from the NaCo HCI observations using the `PynPoint` pipeline (Stolker et al. 2019). First, the data were corrected for dark current, flat field, and bad pixels ($4\sigma$-clipping). The images were aligned using cross-correlation and centred by fitting a 2D gaussian function to the PSF. Finally, since the field of view was not allowed to rotate during the very short sequence, we applied Reference Differential Imaging (RDI, Lafreniere et al. 2009) based on Principal Component Analysis (PCA, Amara & Quanz 2012) to remove the central PSF and achieve the highest possible contrast. As a reference star we chose GL 102, observed during the same program as L 363-38 using the same instrument setup (see Sect. 2.3). The reference star data were calibrated in the same way as those of L 363-38 and were then used to build a Principal Component library able to model the stellar PSF of our target data. No companion is visible in the residuals, and as it usually done in high-contrast imaging we estimated the contrast limits of our data. These were obtained by inserting artificial companions at different separations (between 0.″05 and 3.″0 in steps of 0.″1) and position angles between 0° and 360° in steps of 60° with different brightnesses until a signal-to-noise ratio of 5 is reached. A more detailed description of the procedure used to estimate the contrast limits can be found in (Stolker et al. 2019). We further converted them to mass limits assuming a stellar age of 3-8 Gyr and the AMES-Cond evolutionary models (Baraffe et al. 2003), using the functions provided in the `species` toolkit[8] (Stolker et al. 2020). These are shown in Fig. A.1 (right).

Similar to what presented in Section 3.3, we computed the constraints on companion mass and semi-major axis for possible additional companions orbiting L 363-38 by following the procedure in Boehle et al. (2019), which combines RV and HCI observations. These calculations are based on the ESPRESSO residuals after subtracting the signal from the planet, as well as on the contrast curves computed for the NaCo images. Specifically, the 2d histogram in the upper part of Fig. A.2 shows the percentage of further companions with mass and semi-major axis in a given bin which, if present, could be detected with the current ESPRESSO observations (upper left) and NaCo images (upper right) alone. The lower left plot shows the percentage of companions detected when combining both RV and HCI data, while the lower right plot illustrates the fraction of companions which could be detected in the HCI data but not in the RV ones. Figure 5 shows the same as Fig. A.2 (upper left) but on a different scale, which allows a better comparison with the mass limits in Fig. 4.

In order to compute the constraints, we first simulate Keplerian orbits as described in Section 3.3 but for a grid with $0 < m_p < 100$ M$_J$ and $0 < a < 50$ AU. For each of orbits we then computed the expected RV signal at the epochs of our ESPRESSO observations, and the projected separation at the epoch of the NaCo observation. If the maximum RV difference measured across the observations (i.e. RV$_{max}$ - RV$_{min}$) is higher than 5 times the standard deviation of the measured RVs, then we considered the planet signal as detectable. On the other hand, a companion is considered as detected in the HCI data if its mass is higher than the mass limit in Fig. A.1 (right) at the predicted separation. The completeness for each combination of mass and semi-major axis was then determined from the percentage of detectable planets in the corresponding bin.

As expected, RV is mostly sensitive to planets with high mass and/or small semi-major axis, while HCI is more sensitive at larger separations. Because of the short exposure time (160 seconds) and the available band (NB 1.64, $\lambda_0 = 1.64 \mu m$), the NaCo data do not allow to constrain low-mass planets as it is the case for the ESPRESSO data, but rather only companions in the brown dwarf and stellar regime. New HCI data, e.g. from the Spectro-Polarimetric High-contrast Exoplanet REsearch instrument (SPHERE, Beuzit et al. 2019) or from the upcoming Enhanced Resolution Imager and Spectrograph (ERIS, Davies et al. 2018), are therefore crucial to fully exploit the potential of combining RV and HCI data to better constrain the architecture around nearby stars even in the case of nondetections.

---

[8] `https://species.readthedocs.io/en/latest/index.html`

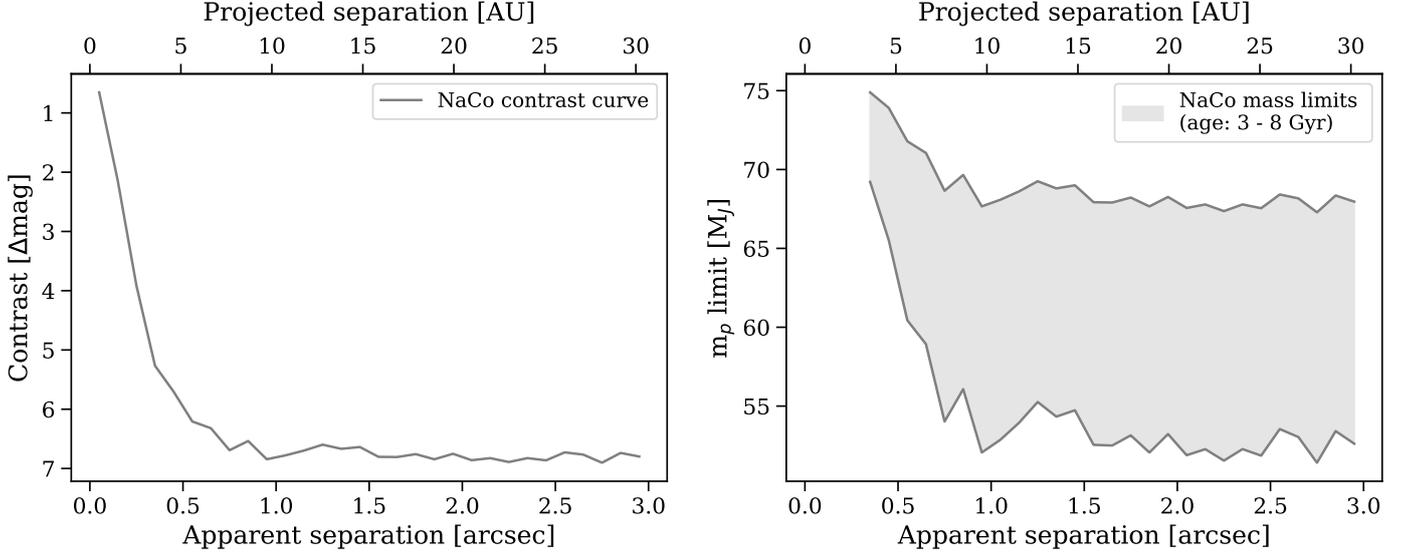

**Fig. A.1.** Limiting contrast curve in the NaCo NB 1.64 filter. *Right*: Limiting companion mass curve computed from the NaCo contrast curve using the AMES-Cond evolutionary models (Baraffe et al. 2003). The shadowed region indicates the spread to due the uncertainty in the stellar age.

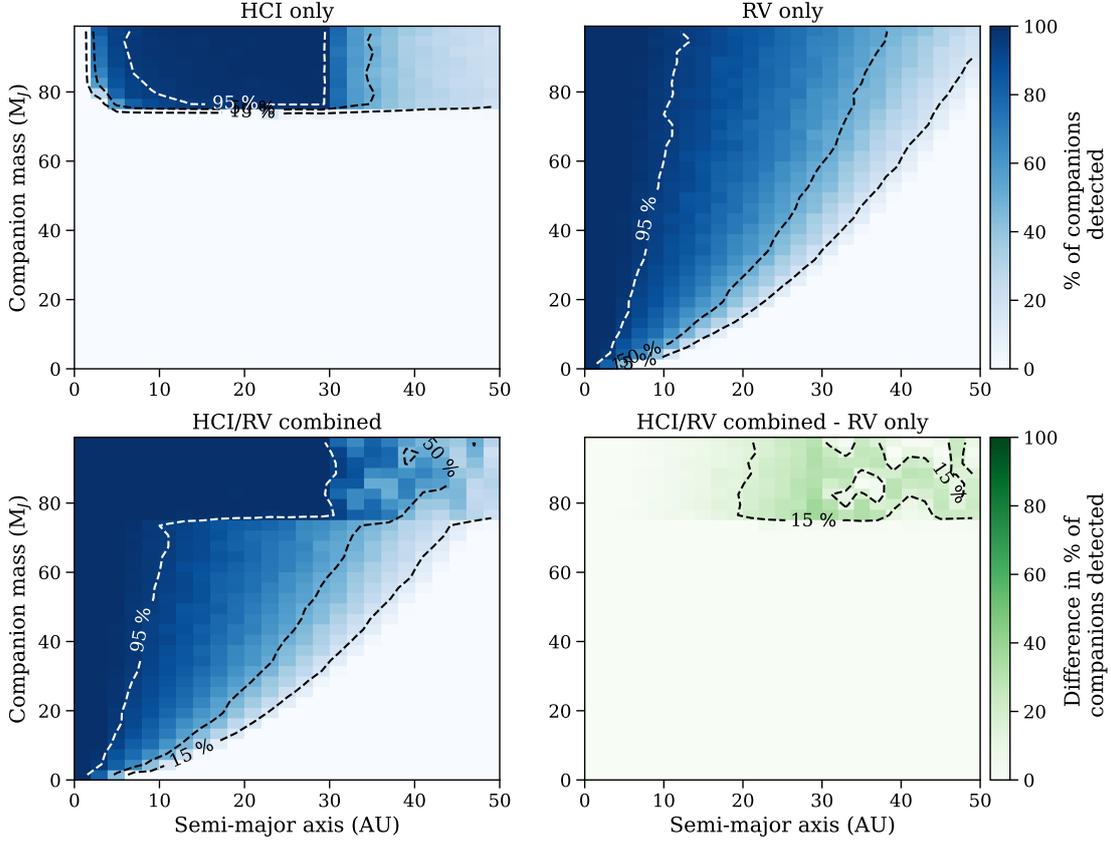

**Fig. A.2.** Constraints on companion mass and semi-major axis for planets orbiting around L 363-38, based on the ESPRESSO residuals after subtracting the signal from the planet and on the contrast curve computed for the NaCo observations. The 2d histograms show the percentage of companions with mass and semi-major axis in a given bin which would be detected, if present, with the current ESPRESSO observations (upper left) and NaCo images (upper right). The percentage of companions detected in at least one of the datasets and in the HCI data only are shown in the lower left and lower right panels, respectively.